\documentclass[prl,twocolumn,amsmath,amssymb,superscriptaddress,showpacs]{revtex4}
\usepackage{graphics}
\usepackage{epsfig}
\usepackage{bbm}
\usepackage[latin1]{inputenc} 

\newcommand{\be}{\begin{equation}}
\newcommand{\ee}{\end{equation}}
\newcommand{\eea}{\end{eqnarray}}
\newcommand{\bea}{\begin{eqnarray}}

\newcommand{\exs}[1]{\ensuremath{\langle{#1}\rangle}}
\newcommand{\mean}[1]{\ensuremath{\langle{#1}\rangle}}

\newcommand{\eins}{\ensuremath{\mathbbm 1}}
\newcommand{\qed}{\ensuremath{\hfill \Box}}
\newcommand{\WW}{\ensuremath{\mathcal{W}}}

\newcommand{\ketbra}[1]{\ensuremath{| #1 \rangle \langle #1 |}}

\newcommand{\ket}[1]{\ensuremath{|#1\rangle}}

\newcommand{\bra}[1]{\ensuremath{\langle#1|}}

\newcommand{\braket}[2]{\ensuremath{\langle #1|#2\rangle}}

\newcommand{\kommentar}[1]{}

\begin{document}
\title{Detecting Genuine Multipartite Entanglement
with Two Local Measurements}
\date{\today}
\begin{abstract}
We present entanglement witness operators for detecting {\it
genuine} multipartite entanglement. These witnesses are robust
against noise and require only two {\it local} measurement
settings when used in an experiment, independent from the number
of qubits. This allows detection of entanglement for an increasing
number of parties without a corresponding increase in effort. The
witnesses presented detect states close to GHZ, cluster and graph
states. Connections to Bell inequalities are also discussed.
\end{abstract}

\author{G\'eza T\'oth}
\email{toth@alumni.nd.edu} \affiliation{Max-Planck-Institut f\"ur
Quantenoptik, Hans-Kopfermann-Stra{\ss}e 1, D-85748 Garching,
Germany,}

\author{Otfried G\"uhne}
\email{otfried.guehne@uibk.ac.at} \affiliation{Institut f\"ur
Theoretische Physik, Universit\"at Hannover, Appelstra{\ss}e 2,
D-30167 Hannover, Germany,}

\affiliation{Institut f\"ur Quantenoptik und Quanteninformation,
\"Osterreichische Akademie der Wissenschaften,
A-6020 Innsbruck, Austria}

\pacs{03.67.Mn, 03.65.Ud, 03.67.-a}

\maketitle


Entanglement lies at the heart of quantum mechanics
and plays an important role in quantum information theory
\cite{BE00}. While bipartite entanglement is well understood,
multi-party entanglement is still under intensive research.
It was soon realized that it is not an extension of the
bipartite case and several new phenomena arise.
For instance, for three qubits there are two different
classes of true many-body entanglement
\cite{DV00}.
Moreover, multi-qubit states can contradict local realistic
classical models in a new and stronger way \cite{GH90}.
These phenomena can be used to implement novel quantum
information processing tasks such as error correction \cite{G96},
fault-tolerant quantum computation, cryptographic protocols such
as secret sharing \cite{CG99}, measurement-based quantum
computation \cite{RB03} and open-destination teleportation
\cite{ZC04}.

With the rapid development of quantum control it is now  possible
to study experimentally the entanglement of many
qubits using photons \cite{ZC04,PB00,ZY03,BE03}, trapped
ions \cite{SK00}, or cold atoms on an optical lattice \cite{MG03B}.
In these experiments it is not sufficient to claim that
"the state is entangled".
A multi-qubit experiment is meaningful and presents
something qualitatively new
only if provably more than two qubits are entangled.
While  lot of thought has been given to detecting entanglement in general
\cite{B64,M90,GB98,HH90},  detection of
{\it genuine} multi-qubit entanglement has
only a limited literature \cite{ZY03,GB98,BE03,GH03}.
Existing methods need an experimental time growing {\it
exponentially} with the number of qubits, making multi-qubit
entanglement detection impossible even for modest size
systems.

We will show, it is still possible to
decide whether a state is multi-qubit entangled
without the need for exponentially growing
resources, using only {\it local}
measurements. This is unexpected since
the property to be detected is {\it nonlocal} over
increasing number of qubits.
Our method can readily be used in any future experiment
preparing GHZ (Greenberger-Horne-Zeilinger) and cluster states
\cite{BR03,RB03}. They both play a central role in the
quantum algorithms mentioned before.
GHZ states, as maximally entangled multi-qubit states,
are intensively studied
\cite{B64,M90,GB98} and
have been realized in numerous experiments
\cite{PB00,ZY03,SK00}. Cluster states can
easily be created in a spin chain with Ising-type interaction
\cite{BR03} and have been realized in optical lattices of
two-state atoms \cite{MG03B}. Remarkably, their entanglement is more
persistent to noise than that of a GHZ state \cite{BR03}.

\begin{figure}
\centerline{\epsfxsize=3in \epsffile{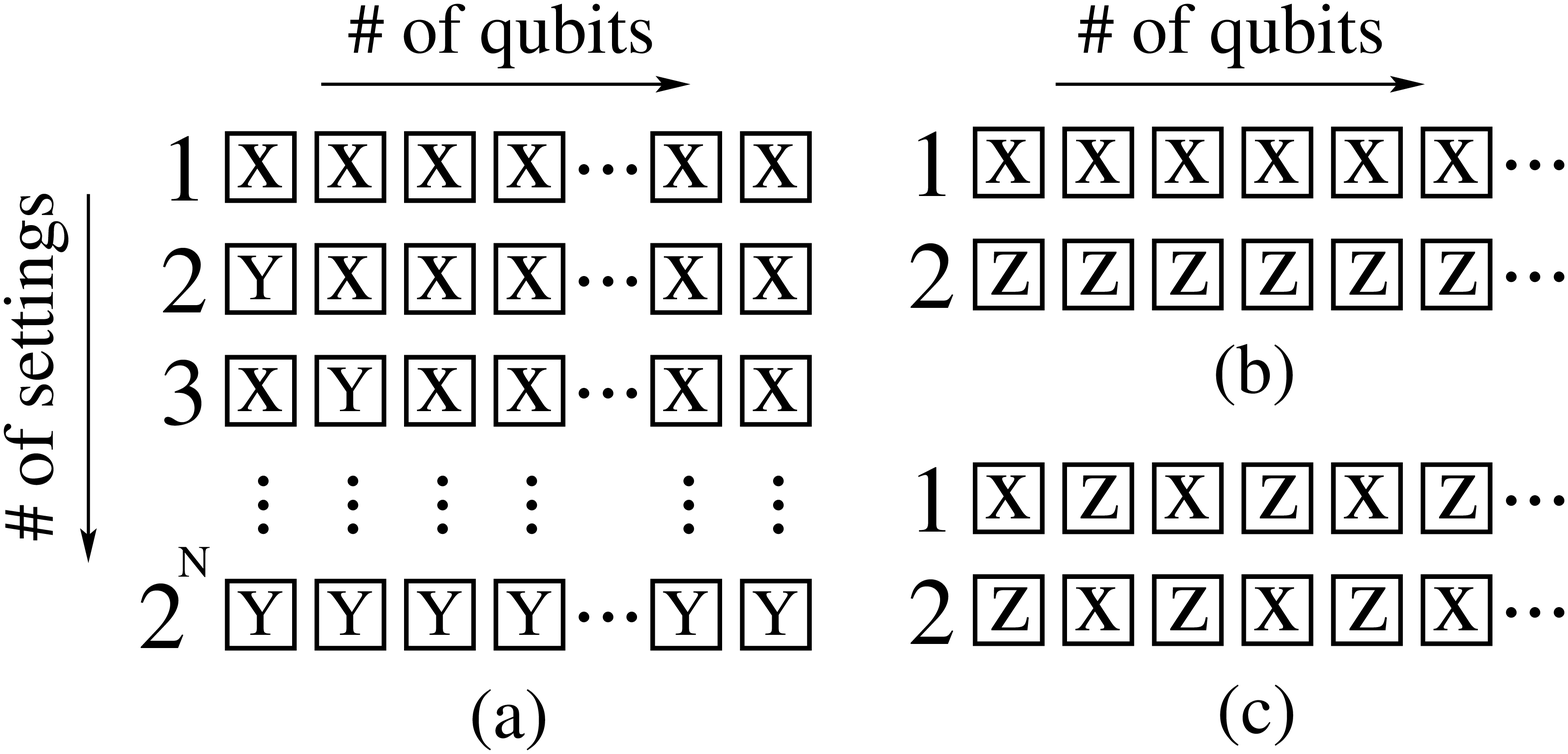} }

\caption{
  (a) Measurement settings needed
  for detecting genuine multi-qubit entanglement close to GHZ states
  with Bell inequalities.
  For each qubit the measured spin component is indicated.
  (b) Settings needed for the approach presented
  in this paper for detecting entangled
  states close to GHZ states and (c) cluster states.}
\label{fig_settings}
\end{figure}

A usual approach for detecting entanglement is using
Bell inequalities \cite{B64, M90, GB98}.
These indicate the violation of local realism,
a notion independent from quantum physics.
When applied to detect quantum entanglement,
they detect usually
any ({\it i.e.}, also  partial or biseparable \cite{BISEP})
entanglement \cite{GB98}.
For $N$ qubits Bell inequalities
typically need the measurement of two variables
at each qubit. Thus, as shown in Fig.~\ref{fig_settings}(a),
the number of local measurement settings
needed increases exponentially with $N$. Here, a measurement
setting means a simultaneous measurement of single qubit operators
$\{O^{(k)}\}_{k=1}^N$ at sites $k=1,2,...,N$ {\it in parallel}.

Another approach for detecting multipartite entanglement is using
entanglement witnesses \cite{HH90}. These are observables which
have a positive or zero expectation value for all separable
states, thus a negative expectation value signals the presence of
entanglement. In a typical experiment one aims to prepare a pure
state, $\ket{\Psi}$, and would like to detect it as true
multipartite entangled. While the preparation is never perfect, it
is still expected that the prepared mixed state is in the
proximity of $\ket{\Psi}$. The usual way to construct entanglement
witnesses using the knowledge of this state is \be
\tilde{\WW}=\tilde{c} \eins-\ketbra{\Psi}. \ee Here $\tilde{c}$ is
the smallest constant such that for every product state
$Tr(\varrho\tilde{\WW}) \ge 0$. In order to measure the witness
$\tilde{\WW}$ in an experiment, it must be decomposed into a sum
of locally measurable operators \cite{exdet}. The number of local
measurements in these decompositions seems to increase
exponentially with the number of qubits \cite{GH03,BE03}.

In this paper we propose to construct witnesses for $N$-qubit
states of the form \be \WW=c_0\eins-\sum_{k} c_k S_k,
\label{wstabil} \ee where the $c_k$'s are constants and the $S_k$
operators stabilize the state $\ket{\Psi}$ \cite{G96} \be S_k
\ket{\Psi} = \ket{\Psi}. \label{stabil} \ee For certain class of
states, {\it i.e.}, for GHZ and cluster states \cite{BR03} the
$S_k$'s can be chosen locally measurable: they are the tensor
products of Pauli spin matrices. It will turn out that for
measuring our {\it stabilizer witnesses}, only two  local
measurement settings are required, independently of the number of
qubits.

Let us shortly explain what we understand by such a local
measurement setting \cite{exdet}. Measuring a local setting
$\{O^{(k)}\}_{k=1}^N$ consists of performing simultaneously the
von Neumann measurements $O^{(k)}$ on the corresponding parties.
After repeating the measurements several times,
the coincidence probabilities for the outcomes are collected.
Given these probabilities it is possible to compute all two-point
correlations $\exs{O^{(k)} O^{(l)}}$, three-point correlations
$\exs{O^{(k)} O^{(l)} O^{(m)}}$, etc. Since  all these correlation
terms can be measured with one setting, the number of settings
determines the experimental effort rather than the number of
measured correlation terms in Eq.~(\ref{wstabil}). For detecting
entanglement at least two settings are needed since the
coincidence probabilities obtained from a single setting can
always be mimicked by a separable state.

In order to demonstrate the power of our approach with an example,
let us write down an entanglement witness (discussed later in
detail) which detects {\it genuine} three-qubit entanglement
around the three-qubit GHZ state $\ket{GHZ_3}=
(\ket{000}+\ket{111})/\sqrt{2}:$
\begin{eqnarray}
\WW_{GHZ_3}&:=& \frac{3}{2}\eins
-\sigma_x^{(1)}\sigma_x^{(2)} \sigma_x^{(3)}\nonumber\\
&-&\frac{1}{2}\big[\sigma_z^{(1)}\sigma_z^{(2)}
+\sigma_z^{(2)}\sigma_z^{(3)} +\sigma_z^{(1)}\sigma_z^{(3)}\big].
\label{ghz3}
\end{eqnarray}
This witness requires the measurement of the
$\{\sigma_x^{(1)},\sigma_x^{(2)},\sigma_x^{(3)}\}$ and the
$\{\sigma_z^{(1)},\sigma_z^{(2)},\sigma_z^{(3)}\}$ settings. The
projector based witness $\WW_{GHZ_3}= \eins/2 - \ketbra{GHZ}$
requires four measurement settings \cite{GH03}.

After showing the previous example, we present
a witness detecting entangled states close to an
$N$-qubit GHZ state, $\ket{GHZ_N}=(\ket{0}^{\otimes
N}+\ket{1}^{\otimes N})/\sqrt{2}$. Its stabilizing operators are
\bea S_1^{(GHZ_N)}&:=& \prod_{k=1}^N \sigma_x^{(k)},
\nonumber\\
S_k^{(GHZ_N)}&:=&\sigma_z^{(k-1)} \sigma_z^{(k)} \mbox{ for }
k=2,3,...,N. \label{eigenGHZ} \eea Using these stabilizing
operators, Eq.~(\ref{stabil}) defines uniquely the GHZ state.
Latter is stabilized not only by $S_k^{(GHZ_N)}$'s but also by
their products. These operators form a group called {\it
stabilizer} \cite{G96}, and $S_k^{(GHZ_N)}$'s are the generators
of this group. Allowing both $+1$ and $-1$ eigenvalues in Eq.
(\ref{stabil}), $2^N$ $N$-qubit states can be defined which are
orthogonal to each other and form a complete basis. We will refer
to this as the {\it GHZ state basis}. All the elements of the
stabilizer are diagonal in this basis.

{\bf Theorem 1.} The following entanglement witness detects
genuine $N$-qubit entanglement for states close to an $N$-qubit
GHZ state: \be \WW_{GHZ_N} :=
3\eins-2\bigg[\frac{S_1^{(GHZ_N)}+\eins}{2}
+\prod_{k=2}^N\frac{S_k^{(GHZ_N)}+\eins}{2}\bigg].
\label{wnoisetol} \ee Another witness for this task is given by
\be \WW_{GHZ_N}':= (N-1) \eins - \sum_{k=1}^{N} S_k^{(GHZ_N)}. \ee

{\it Proof.} First, we need to know that
$\tilde{\WW}_{GHZ_N}=\eins/2-\ket{GHZ_N}\bra{GHZ_N}$ detects
genuine $N$-qubit entanglement. This
follows from the methods presented in Ref.~\cite{BE03}. We will now show
that the witness $\tilde{\WW}_{GHZ_N}$ is finer then the witness
${\WW}_{GHZ_N},$ {\it i.e.}, that for all states with $Tr(\varrho
{\WW}_{GHZ_N})<0$ also $Tr(\varrho \tilde{\WW}_{GHZ_N})<0$ holds
\cite{WITNESS1}.
For that we have to show that
$\WW_{GHZ_N}-\alpha\tilde{\WW}_{GHZ_N}\ge 0$ where $\alpha$
is some positive constant.
Then for any state $\varrho$ detected by
$\WW_{GHZ_N}$ we have
$\alpha Tr(\varrho \tilde{\WW}_{GHZ_N})\le Tr(\varrho{\WW}_{GHZ_N})<0$
thus the state is also detected by $\tilde{\WW}_{GHZ_N}$.
This implies that ${\WW}_{GHZ_N}$ is also a
multi-qubit witness.
Let us now look at the observable
$X:=\WW_{GHZ_N}-2\tilde{\WW}_{GHZ_N}$ and show  that $X \ge 0.$
We can express $X$ in the GHZ state basis. Since ${\WW}_{GHZ_N}$
as well as $\tilde{\WW}_{GHZ_N}$ are diagonal in this basis, $X$
is also diagonal. By direct calculation it is straightforward to
check that the entries on the diagonal are all non-negative, which
proves our claim. For the other witness one can show  similarly
that $\WW_{GHZ_N}'-2\tilde{\WW}_{GHZ_N} \ge 0$
$\qed$

The main advantage of the witnesses ${\WW}_{GHZ_N}$ and
${\WW}_{GHZ_N}'$ in comparison with  $\tilde{\WW}_{GHZ_N}$ lies in
the fact that for implementing them {\it only two measurement
settings} are needed as shown in Fig.~\ref{fig_settings}(b). From
the first setting $\exs{S_1^{(GHZ_N)}}$ can be obtained, from the
second one $\exs{S_k^{(GHZ_N)}}$ for $k=2,3,...,N.$ The form of
${\WW}_{GHZ_N}$ can be intuitively understood as follows. The
first term in the square bracket is a projector to the subspace
where $\exs{S_1^{(GHZ_N)}}=+1$. The second one is a projector to
subspace where $\exs{S_k^{(GHZ_N)}}=+1$ for all
$k\in\{2,3,...,N\}$. Clearly only a GHZ state gives $+1$ for both
projectors. The witness $\WW_{GHZ_N}$ can be proven to be optimal
from the point of view of noise tolerance among stabilizer
witnesses using two measurement settings and 
having the property $\WW_{GHZ_N}-2\tilde{\WW}_{GHZ_N}\ge 0$ \cite{GT04}.

For practical purposes it is important to know how large
neighborhood of the GHZ state is detected by our witnesses. This
is usually characterized by the robustness to noise. The witness
${\WW}_{GHZ_N}$ is very robust: It detects a state of the form
$\varrho(p)=p_{noise} \eins / 2^N + (1-p_{noise}) \ketbra{GHZ_N}$
for $p_{noise}<1/(3-4/2^N)$ as true multipartite entangled thus it
tolerates at least $33\%$ noise, independent from the number of
qubits. For $N=3$ the witness from Eq.~(\ref{wnoisetol}) was
already given in Eq.~(\ref{ghz3}) and tolerates noise up to
$p_{noise} < 0.4$. The witness ${\WW}_{GHZ_N}'$, having  the
minimal $N$ stabilizing terms, is not so robust: It tolerates
noise for $p_{noise} < 1/N.$

Other novel witnesses can be obtained by including further terms
of the stabilizer and using more than two measurement settings.
For instance, following the lines of the previous paragraphs it
can be proved that the observable $\WW_{GHZ_3}'':= 2\eins -
S_1^{(GHZ_3)}[\eins+ S_2^{(GHZ_3)}][\eins+S_3^{(GHZ_3)}] = 2\eins
+ \sigma_y^{(1)}\sigma_y^{(2)}\sigma_x^{(3)} +
\sigma_x^{(1)}\sigma_y^{(2)}\sigma_y^{(3)} +
\sigma_y^{(1)}\sigma_x^{(2)}\sigma_y^{(3)} -
\sigma_x^{(1)}\sigma_x^{(2)}\sigma_x^{(3)} $ detects genuine
three-party entanglement if $p_{noise}<1/2$. It is very remarkable
that witness $\WW_{GHZ_3}''$  is equivalent to Mermin's inequality
\cite{M90} for detecting violation of local realism. However,
Mermin's inequality in the form from above is normally used to
detect some, not necessarily genuine multipartite, entanglement.
From our witness it follows that it  detects indeed only genuine
multipartite entanglement \cite{MERMINMULTIQUBIT}. For $N>3$
Mermin's inequality contains also only stabilizing terms.
Including even more terms from the stabilizer one can even
construct the projector-based witness \cite{STABIL}.

Let us continue our discussion by presenting a witness detecting
entangled states close to cluster states. An $N$-qubit cluster
state, $\ket{C_N}$, can be created starting from the state
$\ket{1111 ...}_x$ by applying the Ising chain-type dynamics
$U_{cl}=\exp\big[i\frac{\pi}{4}\sum_k(1-\sigma_z^{(k)})(1-\sigma_z^{(k+1)})\big].$
The stabilizing operators used for constructing our witnesses are
\bea S_1^{(C_N)}&:=&\sigma_x^{(1)} \sigma_z^{(2)},
\nonumber\\
S_k^{(C_N)}&:=&\sigma_z^{(k-1)} \sigma_x^{(k)} \sigma_z^{(k+1)}
\mbox{ for } k=2,3,...,N-1,
\nonumber\\
S_N^{(C_N)}&:=&\sigma_z^{(N-1)} \sigma_x^{(N)}. \label{eigenC}
\eea The results for cluster states are analogous to the case of
the GHZ state:

{\bf Theorem 2.} The following witnesses detect genuine $N$-party
entanglement close to a cluster state \bea \WW_{C_N} &:=& 3\eins-
2 \bigg[\prod_{\text{even k}}\frac{S_k^{(C_N)}+\eins}{2}+
\prod_{\text{odd k}}\frac{S_k^{(C_N)}+\eins}{2}\bigg],
\nonumber \\
\WW_{C_N}'&:= &(N-1) \eins - \sum_{k=1}^{N} S_k^{(C_N)}.
\label{CN} \eea

{\it Proof.} In order to show that these observables are
witnesses, we first show that \be
\tilde{\WW}_{C_N}:=\frac{1}{2}\eins-\ketbra{C_N}. \ee is a
witness. To do this we have to show that for all pure biseparable
states $\ket{\phi}$ the bound $|\braket{\phi}{C_N}| \leq
1/\sqrt{2}$ holds. This is equivalent to showing that the Schmidt
coefficients do not exceed $1/\sqrt{2}$ when making a Schmidt
decomposition of $\ket{C_N}$ with respect to an arbitrary
bipartite splitting, since they bound the overlap with the
biseparable states \cite{BE03}. It is known that one can produce a
singlet between an arbitrary pair of qubits from a cluster state
by local operations and classical communication \cite{BR03}. For a
singlet both Schmidt coefficients are $1/\sqrt{2}.$ Furthermore,
it is known that the largest Schmidt coefficient cannot decrease
\cite{N99} under these operations. This proves our claim. Knowing
that $\tilde{\WW}_{C_N}$ is a witness, one can show as in the GHZ
case that $\WW_{C_N}$ and  $\WW_{C_N}'$ are also witnesses. $\qed$

The stabilizing operators in the expression given for $\WW_{C_N}$
are again grouped into two terms corresponding to the two
settings shown in Fig.~\ref{fig_settings}(c). The witness
$\WW_{C_N}$ tolerates mixing with noise if
$p_{noise}<1/(4-4/2^{\frac{N}{2}})$ for even $N$ (respectively,
$p_{noise}<1/[4-2(1/2^{\frac{N+1}{2}}+1/2^{\frac{N-1}{2}})]$ for
odd $N$). Thus, for any number of qubits at least $25 \%$ noise
are tolerated. Alternatively, $\tilde{\WW}_{C_N}$ can
also be decomposed into local terms following
Refs.~\cite{BE03,GH03}. The noise tolerance is at least $50\%$
even for large $N,$ however, more than the two settings are
necessary.

Up to now, we presented witnesses detecting {\it only} genuine
$N$-qubit entanglement. If the noise is large, there might be no
true $N$-party entanglement in the system. In this case {\it some}
entanglement can still be detected with the two measurement
settings from above, although it may not be multipartite
entanglement. Similarly to Ref.~\cite{T03}, the following
necessary conditions for full separability can be constructed for
GHZ and cluster states \bea
\exs{S_1^{(GHZ_N)}}+\exs{S_m^{(GHZ_N)}}&\le& 1 \;\; {\rm for}\;\;
N \ge m \ge 2
\label{SepGHZ},\\
\exs{S_k^{(C_N)}}+\exs{S_{k+1}^{(C_N)}}&\le& 1 \;\;{\rm for }\;\;
N-1 \ge k \ge 1.\;\;\;\;\;\;\; \label{SepC} \eea These conditions
detect entanglement after mixing with noise if $p_{noise}<1/2$ and
they both need only two measurement settings. The proofs are given
in the Appendix.

The previous results can straightforwardly be generalized for
graph states \cite{DA03}. These states are defined by a graph of
$N$ vertices. Edges of this graph are described by the adjacency
matrix $\Gamma$. $\Gamma_{kl}=1$ $(0)$ if the vertices $k$ and $l$
are connected (not connected). An $N$ qubit state is defined as an
eigenstate with eigenvalue $1$ of the stabilizing operators
$S_k^{(G_N)}:=\sigma_x^{(k)} \prod_{l\ne k}
(\sigma_z^{(l)})^{\Gamma_{kl}}$.  Physically, $\Gamma_{kl}=1$
$(0)$ means that spins $k$ and $l$ interact (does not interact) by
an Ising-type interaction. We assume, that the graph cannot be
partitioned into two separate subgraphs, since then the graph
state would be biseparable.

A witness detecting genuine $N$-party entanglement can be defined
as $\WW_{G_N} := (N-1)\eins - \sum_k S_k^{(G_N)}.$ The proof is
essentially the same as before. It must be used that one can
produce from a graph state by local means a singlet between an
arbitrary pair of qubits \cite{eisert04}. For two-colorable graphs
only two settings are needed for measuring $\WW_{G_N}$
\cite{BRIEGEL}. The maximum number of settings required is $N$,
reached for example by the state corresponding to the complete
graph. A necessary condition for separability can be given as
$\exs{S_k^{(G_N)}}+\exs{S_m^{(G_N)}}\le 1$ where spin $(k)$ and
$(m)$ are neighbors.

In summary, based on the stabilizer theory we constructed
entanglement witnesses with simple local decomposition for GHZ,
cluster and graph states. Our approach is optimal from the point
of view of the duration of an experimental implementation since
only two local measurement settings are needed independent from the
number of qubits. We found that some Bell inequalities (when used for
entanglement detection) and the
projector based witnesses are in fact also stabilizer witnesses.

We thank M.~Aspelmeyer, H.J.~Briegel, D.~Bru{\ss}, {\v
C}.~Brukner, J.I.~Cirac, T.~Cubitt, M.~Dreher, J.~Eisert,
J.J.~Garc\'{\i}a-Ripoll, P.~Hyllus, M.~Lewenstein, A.~Sanpera,
V.~Scarani, M.M.~Wolf, and  M.~\.Zukowski for useful discussions.
We also acknowledge the support of the DFG (Graduierten\-kolleg
282), the European Union (Grant No. MEIF-CT-2003-500183), the EU
projects RESQ and QUPRODIS, and the Kom\-pe\-tenz\-netz\-werk
Quan\-ten\-in\-for\-ma\-tions\-ver\-ar\-bei\-tung der
Bayerisch\-en Staats\-re\-gie\-rung.

{\it Appendix: Proof of Eqs. (\ref{SepGHZ},\ref{SepC}).} Using the
Cauchy-Schwarz inequality and the fact that
$\exs{\sigma_x^{(k)}}^2+\exs{\sigma_z^{(k)}}^2\le 1$ we obtain for
product states $\exs{S_1^{(GHZ_N)}}+ \exs{S_m^{(GHZ_N)}}
\le|\exs{\sigma_x^{(m-1)}}|\cdot|\exs{\sigma_x^{(m)}}|+
|\exs{\sigma_z^{(m-1)}}|\cdot|\exs{\sigma_z^{(m)}}|\le 1$ for
$m=2,3,...,N$. Due to linearity, this bound is also valid for full
separable states. For the second inequality, we have
$\exs{S_k^{(C_N)}+S_{k+1}^{(C_N)}} =
\exs{\sigma_z^{(k-1)}}\exs{\sigma_x^{(k)}}\exs{\sigma_z^{(k+1)}} +
\exs{\sigma_z^{(k)}}\exs{\sigma_x^{(k+1)}}\exs{\sigma_z^{(k+2)}}
\le
  |\exs{\sigma_x^{(k)}}|\cdot|\exs{\sigma_z^{(k+1)}}|
+ |\exs{\sigma_z^{(k)}}|\cdot|\exs{\sigma_x^{(k+1)}}| \le 1.$
Here, for the end of the chain
$\sigma_z^{(0)}=\sigma_z^{(N+1)}=\eins$ was used. $\qed$



\begin{thebibliography}{99}
\bibitem{BISEP}
Creating pure genuine
multipartite entangled states requires the participation of
all parties. Pure biseparable states, however, can be created such
that two groups of the parties never interact and parties
corresponding to different groups remain unentangled.
A mixed state is called biseparable if it is
the mixture of biseparable pure states. This mixture may contain
states which are biseparable with respect to different partitions.
\bibitem{STABIL}
The projector onto a GHZ state can also be written as
$\ketbra{GHZ_N}=2^{-N}\prod_{k=1}^N({S_k^{(GHZ_N)}+\eins}).$ This
can be seen by expanding both sides in the GHZ state basis.
\end{thebibliography}
\end{document}